\documentclass[12pt]{article}
\usepackage{amsmath,amsthm}
\usepackage{amssymb}
\usepackage{graphicx,psfrag,epsf}
\usepackage{enumerate}
\usepackage{natbib}
\usepackage{bm}
\usepackage{multirow}
\usepackage{caption}
\usepackage{subcaption}
\usepackage{float}
\usepackage[title]{appendix}
\usepackage{soul}
\usepackage{color}
\usepackage[shortlabels]{enumitem}
\setlist[enumerate]{noitemsep}
\usepackage{mathrsfs}

\newtheorem{theorem}{Theorem}

\begin{document}

\title{A Bayesian Nonparametric System Reliability Model which Integrates Multiple Sources of Lifetime Information}

\author{Richard L. Warr\footnote{Corresponding Author: warr@stat.byu.edu} , Jeremy M. Meyer, and Jackson T. Curtis}
         
\maketitle

\begin{abstract}
We present a Bayesian nonparametric system reliability model which scales well and provides a great deal of flexibility in modeling.  The Bayesian approach naturally handles the disparate amounts of component and subsystem data that may exist.  However, traditional Bayesian reliability models are quite computationally complex, relying on MCMC techniques.  Our approach utilizes the conjugate properties of the beta-Stacy process, which is the fundamental building block of our model.  These individual models are linked together using a method of moments estimation approach.  This model is computationally fast, allows for right-censored data, and is used for estimating and predicting system reliability.
\end{abstract}

\noindent{\sc KEY WORDS:  Beta-Stacy Process, Conjugate Models, Dirichlet Process, Method of Moments, Right Censoring}

\section{Introduction}

System reliability models are crucial in determining important quantities, such as system safety, warranty costs, and product lifetimes.  Typical systems have many components with relatively sparse amounts of system data.  However, data for components are usually much more abundant and cheaper to obtain.  A simple model that exclusively considers system data is bound to be inferior to a more sophisticated model which includes additional data.  This work presents a system reliability model with the aim of incorporating as much data as are available, while minimizing assumptions.

To accomplish this goal, we adopt a Bayesian nonparametric model at each step, and present a technique to fuse these models together at each level.  There are significant benefits of this approach.  First, the Bayesian framework provides a natural and disciplined way to incorporate disparate types of data at each level of the process.  Additionally, the nonparametric model reduces the number assumptions needed for the model to hold.  Finally, the specific model we use, as a building block, leverages conjugate properties and eliminates the need for Markov chain Monte Carlo (MCMC) sampling, which can be time consuming to run and tune.  

The basic building block at each stage of our model uses the beta-Stacy process (BSP) which was introduced in \cite{walker-1997-beta}.  This Bayesian nonparametric model is structurally conjugate, in that if a prior is a BSP and if the data are independent and identically distributed (with possible right censoring), then the posterior is also a BSP.  Each component of the system is modeled with a BSP, which are then fused together using a method of moments technique for the BSP.  A contribution of this work is the derivation of the second moment of the discrete BSP, which facilitates our approach.

The remainder of the paper is organized as follows.  First we discuss of the relevant literature of system reliability modeling.  A detailed look at the beta-Stacy process is provided, with the derivation of its second moment function.  With this information in-place, the statistical model is given in detail and some of its properties are explored.  This is followed by a simulation study, which demonstrates the advantages our model and when it is best suited to implement.  Next, the model is applied to an industrial system in which its capabilities are demonstrated.  Finally, we conclude with a discussion of the model's properties and possible future improvements.

\subsection{Literature Review}

In reliability analysis, it is often the goal to estimate an unknown cumulative distribution function (CDF). A key characteristic of reliability data, however, is often the presence of right censored observations. In system reliability one is interested in the measuring the reliability of that system which is composed of smaller components, each with their own CDF describing the probability of failure at or before a given point in time. Let $F(t)$ denote the CDF and $R(t)=1-F(t)$ represent the reliability function.

Much of the early work on Bayesian system reliability involved binomial data (e.g., the components either passed or failed inspection), as in \citet{cole-1975-bayesian}. Other important works include \citet{mastran-1976-incorporating} and \citet{mastran-1978-bayesian}.  However, more modern work forcuses on modeling a system's lifetime.

Traditionally, system lifetime models are approached from a frequentist paradigm.  This is apparent from the sparsity of textbook expositions of Bayesian system reliability.  For example \cite{rausand2003system}, a textbook on system reliability, contains a chapter on Bayesian system reliability.  Also, the textbooks \cite{martz1982, hamada2008bayesian} on Bayesian reliability each contain a chapter on system reliability.  This highlights the relatively modest amount of work that is available in this area.

However, the amount of literature on Bayesian system reliability has been increasing.  An influential paper by \cite{reese-2011-bayesian} proposes a model for assessing the reliability of complex multicomponent systems that utilized Markov Chain Monte Carlo (MCMC) methods to accomplish Bayesian inference. The Bayesian nature of their method easily allows for the pooling of information across similar components, and the incorporation of expert opinion.  A strength of this method is that it allows inference to go both ways; the component data will influence the posterior of the system and the system data to influence the posterior of the components.  A draw back of their approach is there are many parametric assumptions which, if misspecified, could accumulate and result in an unsatisfactory system model.  The main goal of our work is to obtain a system estimate, with little regard for learning more about the components (since we assume ample component data are present). 

\cite{reese-2011-bayesian}'s work is further extended in \cite{guo-2013-bayesian} to binary, lifetime, degradation, and expert opinion data at any level of the system.  Additionally, \cite{jian2018bayesian} and \cite{guo2018system} use data and information from various sources of test data and expert knowledge at subsystem and system levels. They both use Bayesian melding methods to combine information.  They use Markov Chain Monte Carlo (MCMC) and adaptive Sampling Importance Re-sampling (SIR) methods to make inference on the  system reliability.  These important works extend the literature in the area of Bayesian parametric system reliability, however, we diverge and take a nonparametric approach.  

While Bayesian system reliability has seen a relatively small amount of attention, system reliability using Bayesian nonparametric estimation is even more limited.  Here we give a brief introduction to some fundamental Bayesian nonparametric literature.  

If we know little about $F(t)$ (or equivalently $R(t)$), then a nonparametric approach for inference and predictions can be appealing. Within the frequentist framework, this is typically done by estimating $F(t)$ using the empirical distribution function, or in the case of censored data, the Kaplan-Meier estimator \citep{kaplan-1958-nonparametric}. However, in the Bayesian framework a prior process is specified on the space of all distribution functions $\mathscr{F}$, e.g., a Dirichlet Process \citep{ferguson1973bayesian}. 
\citet{susarla-1976-nonparametric} use a Dirichlet process (DP) prior for the unknown distribution function $F(t)$. However, the DP may not provide an adequate model for $F(t)$ in the case of censored lifetime data.  \cite{doksum-1971-tailfree} introduced the \textit{neutral to the right process} (NRP), a special class of random distribution functions. \citet{ferguson-1979-bayesian} extend their results for censored data.  An important feature of NRPs is: given an NRP prior and data (which may be randomly right censored) the posterior is also a NRP. In this sense, a NRP prior is structurally conjugate because the prior and posterior processes are of the same type.   

\cite{walker-1997-beta} introduced the Beta-Stacy process (BSP) prior, a neutral to the right process prior that is structurally and parametrically conjugate. The authors also show the BSP is a generalization of the Dirichlet process.  As a result, given right-censored observations and a DP prior for $F(t)$, the posterior of $F(t)$ is a Beta-Stacy process. Additionally, when no prior information is given (or when the prior process has a precision of zero), the centering measure of the BSP posterior is equivalent to the Kaplan-Meier estimator. Thus, the BSP provides a convenient prior for the space of cumulative distribution functions when right censoring is present.

\cite{warr-2013-bayesian} use the \textit{Dirichlet process} (DP) to make a nonparametric assessment of the reliability of multi-component systems. A drawback of this approach, however, is that it assumes the data are not censored. In this paper we take a similar approach to modeling system reliability.  Our work extends their methodology by allowing for right-censored data.  This is accomplished by replacing the Dirichlet process with the more general \textit{Beta-Stacy process}.  Additionally, their work was somewhat limited by an ad hoc method of uncertainty propagation.  In our work we solve this more generally, and preserve the uncertainty at each level of the system. 

We recognize there is much more literature system reliability assessment from both frequentist and Bayesian perspectives than we can cover here.  We refer the reader to \cite{reese-2011-bayesian} and \cite{guo-2013-bayesian} for a more thorough overview of this literature.

\section{The Beta-Stacy Process}
In this section we discuss the salient aspects of the Beta-Stacy process (BSP) in our modeling approach.  In our use of the BSP we make a few minor modifications to that presented in \cite{walker-1997-beta} to suite our purposes, the details are described in Appendix \ref{apx:changes}.  The BSP is a generalization of the Dirichlet Process \citep{ferguson1973bayesian}. It is one model in a larger class of neutral to the right processes.  BSPs can be used to model an unknown CDF $F(t)$. 
As data are collected the processes converges weakly to the true distribution.  Thus, from a Bayesian perspective one's belief/knowledge of $F(t)$ is considered random, as data are collected about $F(t)$, one's belief  about $F(t)$ should approach the actual function.    

A BSP is defined by two functional parameters.  The first is denoted as $\alpha(t)$; which we call the precision parameter.  The precision parameter must be positive; the larger the precision, the more certain the belief of $F(t)$. Due to the nature of right censoring, the precision function is left continuous.
One distinguishing difference between the BSP and the DP is that the BSP's precision parameter varies across $t$, whereas the precision parameter of the DP is a constant for all $t$.  This variable precision parameter in the BSP provides enough flexibility to still retain ``conjugacy," even with right censored data.  The DP is only considered to be conjugate with fully observed data (i.e., for data where the exact failure time is known).  

The second parameter is denoted by $G(t)$, which we refer to as the centering measure. Because it is a CDF, the centering measure is right continuous. The centering measure is the expected value of the BSP, as stated in the following theorem.  
\begin{theorem}
\label{thm:mean}
For a given $t$, if $F(t) \sim BSP(\alpha(t),G(t))$ then $E[F(t)]=G(t)$.
\end{theorem}
The proof of Theorem \ref{thm:mean} for a discrete BSP is contained in Appendix \ref{apx:precision}.
For computational efficiency, in this work we confine $G(t)$ to be the CDF of a discrete random variable.  This does not limit or preclude the true CDF, $F(t)$, from being a continuous, however, our estimate of it will be discrete.  As the amount of data increases this becomes a minor technical point.

For the sake of simplicity we avoid discussing how the BSP is formally defined as a L\'evy process.  However, for our approach it is necessary to know the first two moments of $F(t) \sim BSP(\alpha(t), G(t))$.  The calculations are greatly simplified with the restriction that $G(t)$ is the CDF of a positive discrete random variable.  The first moment is defined in Theorem \ref{thm:mean} and is stated in \cite{walker-1997-beta}.  We derive the second moment of a BSP using a property of the jumps in the L\'evy process.  The distribution of the jumps can be transformed to a beta distribution and the jumps of the L\'evy process are related back to $F(t)$ (the CDF of a random measure with a BSP distribution).  Define $n$ to be the number of jumps in $G(t)$ and $t_1 < t_2 < \cdots < t_n$ to be all the time increments such that $G$ is discontinuous at each (i.e.,
\begin{equation*}
\lim_{t \uparrow t_k}G(t) \neq \lim_{t \downarrow t_k}G(t)
\end{equation*}
 $ \forall \, k \in \{1,2,\ldots,n\}$).
Additionally, define $t_0 \equiv 0$ and $t_{n+1} \equiv \infty$.  
\begin{theorem}
\label{thm:2ndmoment}
For an arbitrary $t>0$ let $m \in \{0,1,2,\ldots,n\}$ be the unique index such that $t_{m} \leq t < t_{m+1}$ then
\begin{equation*}
E[(F(t))^2] = \left( \prod_{i=1}^{m} \frac{(1-G(t_i))[\alpha(t_i)(1-G(t_i))+1]}{(1-G(t_i-))[\alpha(t_i)(1-G(t_i-))+1]} \right) -1 +2G(t),
\end{equation*}
where 
\begin{equation*} G(t_i-) = \lim_{t \, \uparrow \, {t_i}}G(t).\end{equation*}
\end{theorem}
\noindent We also note that since we are dealing with a discrete process $G(t_i-)=G(t_{i-1})$.  The proof of Theorem \ref{thm:2ndmoment} is also contained in Appendix \ref{apx:precision}.

Now assume some \textit{a priori} information about $F(t)$ is contained in the prior $F(t) \sim BSP(\alpha(t),G(t))$ and data from the random variable $T_i \ {\buildrel \rm \textit{iid} \over \sim} \ F$ are collected. The purpose for using the BSP in this setting is to allow for right censoring, therefore the data vector, $\bm{T}$, may include right censored observations which are denoted in the corresponding vector $\bm{C}$. Let $d$ be the total number of observations in the data.  If $C_1=1$ then $T_1$ is not censored and if $C_1=0$ then $T_1$ is right censored.
Define 
\[M(t) = \sum_{i=1}^{d} \, \text{I}  (T_i \geq t), \text{ (number of units not failed just before time } t \text{) and } \]
\[J(t) = \sum_{i=1}^{d} C_i \, \text{I}(T_i = t), \text{ (number of failures that occurred at time } t \text{).}\]

The posterior $F(t) \,|\,\bm{T}$, as defined in \cite{walker-1997-beta}, is also a BSP. Consider the set of support points where the BSP's posterior centering measure has jumps.  This set is a union of the support point jumps in the BSP's prior centering measure and the set containing the observed data. This set of discontinuous time points in the posterior's center measure can be denoted as $\{t_1 < t_2 < ... < t_{n^{*}} \}$ with cardinality $n^{*}$. Let $t_0 \equiv 0$ and $m^{*} \in \{0,1,2,...,n^{*}\}$ be the index such that $t_{m^{*}} \le t < t_{m^{*}+1}$. Thus, the BSP posterior has the centering measure defined by
\begin{equation}
G^*(t) = 1-\prod_{i=1}^{m^{*}} \left({1-\frac{\alpha(t_i)(G(t_i)-G(t_i-))+J(t_i)}{\alpha(t_i)(1-G(t_i-))+M(t_i)}}\right)
\label{Eq:postbase}
\end{equation}
and precision parameter
\begin{equation}
\alpha^*(t) = \frac{\alpha(t)(1-G(t))+M(t)-J(t)}{1-G^*(t)} .
\label{Eq:postprec}
\end{equation}
Note that when the prior precision is set to zero, the posterior BSP centering measure is closely related to the Kaplan-Meier estimator. However, since the posterior centering measure will be equal to 1, a prior precision of 0 will result in an undefined precision for any point above the maximum data point (call this $T_{max}$). 

A couple of simple examples of finding the posterior of the BSP are shown in Appendix \ref{apx:post}.  We will now use these basic formulas and properties of the BSP to build the proposed system reliability model.

\section{Methodology}

This section describes how the proposed system reliability model is assembled.  
We initially show how to find the first and second moments of two components being merged into a subsystem.  Then the process for finding the new centering measure and precision of that subsystem is explained, which defines a BSP prior for that subsystem.  We then outline the steps to assemble the total system model.  A short discussion on component priors is given, and finally, we conclude with a subsection on the computational aspects of our model.

\begin{figure}[!ht]
        \centering
        \begin{subfigure}[b]{0.43\textwidth}
                \includegraphics[width=\textwidth]{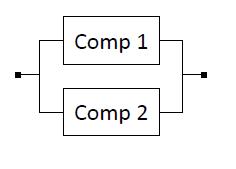}
                \caption{Two components in parallel}
                \label{Fig:parallel}
        \end{subfigure}%
        ~ 
        \begin{subfigure}[b]{0.55\textwidth}
                \includegraphics[width=\textwidth]{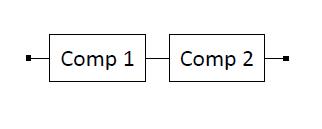}
                \caption{Two components in series}
                \label{Fig:series}
        \end{subfigure}
        \caption{Two component systems}
        \label{Fig:ParandSer}
\end{figure}

The primary step in finding the first and second moments of two combined components is to determine whether they are in \textit{in series} or \textit{in parallel}.  This can readily be determined from a reliability block diagram of the subsystem.  
For two components in parallel, their combined information is contained in the maximum of the two random variables.  In Figure \ref{Fig:parallel} the subsystem CDF for two components in parallel is:
\[ F_S(t) = P(X_1 \leq t \text{ and } X_2 \leq t) = P(X_1 \leq t)P(X_2 \leq t) = F_{X_1}(t)F_{X_2}(t).\]
Basically the two centering measures of the components are multiplied together to obtain the centering measure for the subsystem. Therefore
\begin{align}
\label{eq:1momPara} 
\begin{split}
 G_S(t) \equiv & E[F_S(t)] =  E[F_{X_1}(t)F_{X_2}(t)]= \\ & E[F_{X_1}(t)]E[F_{X_2}(t)]=G_1(t)G_2(t).
 \end{split}
 \end{align}
These formulae are true only if each component is independent of one another, which is a fundamental assumption of our model.
If two components are in series (as in Figure \ref{Fig:series}) the subsystem CDF is:
\begin{align*} F_S(t) &= P(X_1 \leq t \text{ or } X_2 \leq t)\\ &= 1-P(X_1 > t \text{ and } X_2 > t) \\ &=1-P(X_1 > t)P(X_2 > t) \\ &= 1-R_{X_1}(t)R_{X_2}(t). \end{align*}
Again $R(t) \equiv 1-F(t)$, therefore
\begin{align}
\label{eq:1momSeries}
\begin{split}
G_S(t) \equiv E[F_S(t)] &= E[1-R_{X_1}(t)R_{X_2}(t)] \\ &= 1-E[1-F_{X_1}(t)]E[1-F_{X_2}(t)]\\ &= G_1(t)+G_2(t)-G_1(t)G_2(t).
\end{split}
\end{align}

Although here we only discuss two components in series or in parallel the principles are easily extended for many components in any configuration of a reliability block diagram.  Finding the second moment for the subsystem prior is also fairly straightforward.  

For the components in parallel the second moment of the subsystem is:
\begin{equation} \label{eq:2momPara}
E[(F_S(t))^2] = E[(F_{X_1}(t)F_{X_2}(t))^2] = E[(F_{X_1}(t))^2]E[(F_{X_2}(t))^2].
\end{equation}
See Theorem \ref{thm:2ndmoment} to calculate the last two quantities of Equation \ref{eq:2momPara}.
For the components in series, the second moment of the subsystem is:
\begin{align} \label{eq:2momSeries}
\begin{split}
E[(F_S(t))^2] = \ &E[(1-R_{X_1}(t)R_{X_2}(t))^2] \\ = \ & 1-2E[R_{X_1}(t)R_{X_2}(t)]+E[(R_{X_1}(t)R_{X_2}(t))^2]\\
= \ &1-2E[R_{X_1}(t)]E[R_{X_2}(t)]+E[(R_{X_1}(t))^2]E[(R_{X_2}(t))^2]\\
= \ &1-2(1-G_1(t))(1-G_2(t))+(1-2G_1(t)+ \\ \ & E[(F_{X_1}(t))^2])(1-2G_2(t)+ E[(F_{X_2}(t))^2])
\end{split}
\end{align}
The quantities from this last equation are known from the components. Although this procedure may seem mathematically tedious, it is numerically fast.

The previous steps provide the first and second moments for the subsystem prior. The resulting process of this combined component information is not a BSP, but we approximate it with one. We argue that since a BSP provides a flexible nonparametric estimate of a CDF the approximating BSP will also converge to the truth (if the model assumptions hold).   Therefore, using the method of moments we find a BSP with the same first and second moments as the combined components. 

The centering measure for the BSP prior is just the first moment of the BSP. The precision for the prior BSP is a function of the first and second moments.  Referring to Theorem \ref{thm:2ndmoment}, we replace $E[(F(t))^2]$ with the second moment of the combined prior information and $G(t)$ with its first moment. For any $t$ the only unknowns in the equation are the $\alpha(t_i)$. Recall that the precision is constant except at the jumps of $G(t)$, so let $m$ be the $m^{th}$ jump at, or immediately before, time $t$. Thus, $\alpha(t) = \alpha(t_m)$. By incorporating the second moment of the previous jump, we can obtain a recursive formula relating the precision at $t_m$ to the second moment. This formula is stated in the following theorem and details of the derivation are shown in Appendix \ref{apx:precision}.  For simplicity of presentation we define the following:
\begin{align*}
Numerator_1 &= \Big(E\big[\big(F_S(t_{m-1})\big)^2\big]+1-2G_S(t_{m-1})\Big)\big(1-G_S(t_m)\big) \\
Numerator_2 &= \Big(E\big[\big(F_S(t_{m})\big)^2\big]+1-2G_S(t_{m})\Big)\big(1-G_S(t_{m-1})\big) \\
Denominator_1 &= \Big(E\big[\big(F_S(t_{m})\big)^2\big]+1-2G_S(t_{m})\Big)\big(1-G_S(t_{m-1})\big)^2 \\
Denominator_2 &= \Big(E\big[\big(F_S(t_{m-1})\big)^2\big]+1-2G_S(t_{m-1})\Big)\big(1-G_S(t_{m})\big)^2.
\end{align*}
\begin{theorem}
\label{thm:prec}
Given some $t>0$, the corresponding $t_m$,  $E\big[\big(F_S(t_{m-1})\big)^2\big]$, $E\big[\big(F_S(t_{m})\big)^2\big]$, $G_S(t_{m-1})$, and $G_S(t_{m})$ then the precision at $t$ is
\begin{equation}
\alpha(t_m) = \frac{Numerator_1 - Numerator_2}{Denominator_1 - Denominator_2}.
\label{Eq:prec}
\end{equation}
\end{theorem}
For Equation \ref{Eq:prec} we define $E[(F_S(t_{0}))^2] \equiv 0$. Likewise, in rare cases where the method of moments estimator gives negative values for a precision, we also define it to be 0. 

\begin{figure}[!ht]
\begin{center}
\includegraphics[width=3in]{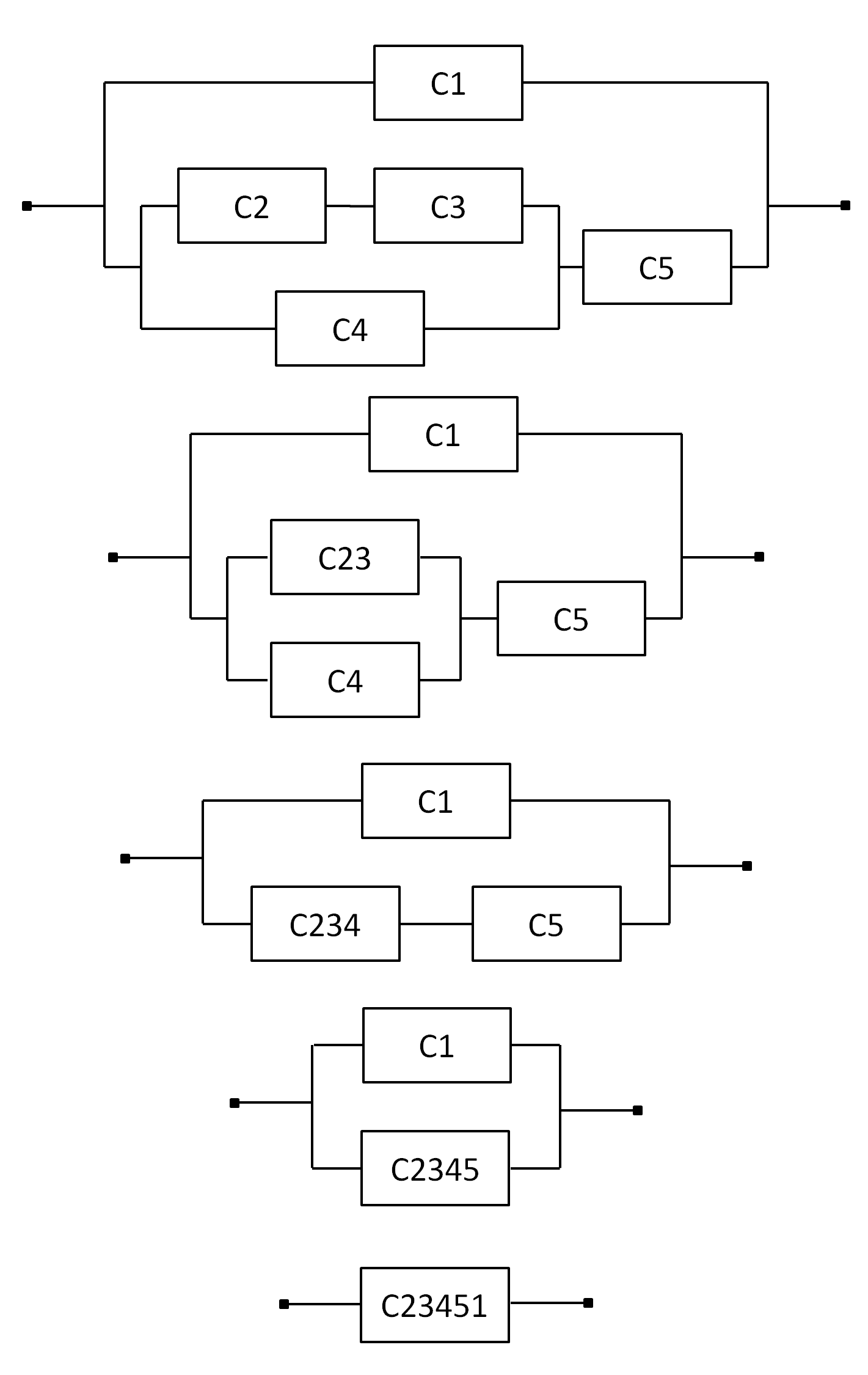}
\caption{A step by step example of how to combine components two at-a-time to reduce a subsystem to a conceptually new component.}
\label{Fig:MultiComps}
\end{center}
\end{figure}

Equations \ref{eq:2momPara}, \ref{eq:2momSeries}, and  \ref{Eq:prec} provide the relationships needed to define the BSP prior for a simple two component subsystem.  Most subsystems are not composed of just two components. Regardless of the subsystem complexity, pairs of components can be combined using the equations above to conceptually model a new component.  This procedure is repeated until the first two moments of the subsystem are known.  For an example of combining components two at-a-time see Figure \ref{Fig:MultiComps}.  At each step in the figure a new component is defined and the subsystem is simplified.  

We have outlined the process to find a BSP to model a subsystem from its components.  Now assign this BSP as the prior for the subsystem, and once subsystem data are available, Equations \ref{Eq:postbase} and \ref{Eq:postprec} are applied to find the posterior for the subsystem.  

Treating a subsystem as a component in a larger subsystem, the process can be iterated as many times as necessary.  We place high importance on computational efficiency and the above procedures are quite fast.  In fact, as the complexity of a system grows, the computation time of the model does not increase at an exponential rate and appear to be smaller than $O(N^2)$. The computational complexity of the model is explored in Section \ref{sec:comp_time}, where more components and data are added to determine the effects on computation time.  

\subsection{The System Model}
The goal of the model is to use the information of the components as a prior for the subsystem.  The basic concept is quite simple and can be recursively applied to a system with many components.  The process begins first at the lowest level; the components are assigned BSP priors, and then combined with their data the posterior of each component is found (which are also BSPs).  Next, combining two at a time, we use a method of moments approach to find a BSP whose moments match the subsystem composed of its components.  This BSP is assigned to be the prior for that subsystem.  Then this subsystem BSP prior is updated with subsystem data to obtain a BSP posterior for the subsystem.  This process is then iterated, treating the subsystems as components at the next level in the system hierarchy.

We propose the following steps for a particular subsystem with $n$ components:  
\begin{enumerate}
\item For each component, let its prior be
\[F_i(t) \sim BSP\left(\alpha_i,G_i(t)\right), \] 
where $G_i(t)$ is a discrete centering measure and $\alpha_i$ is a non-negative constant.  If no prior information exists, set its precision parameter to zero (in which case the choice of $G_i(t)$ is irrelevant). 
\item Next find each component's posterior 
\[F_i(t) \mid \bm{T}_i \sim BSP\left(\alpha^*_i(t),G^*_i(t)\right), \]
using Equations \ref{Eq:postbase} and \ref{Eq:postprec}.
\item Compute the first and second moments from the BSP posterior for each component using the equations in Theorems \ref{thm:mean} and \ref{thm:2ndmoment}.  
\item Calculate the first and second moments of the merged components, two at a time, using Equations \ref{eq:1momPara} and \ref{eq:2momPara} for components in parallel and Equations \ref{eq:1momSeries} and \ref{eq:2momSeries} for components in series.  When there are more than two components, the concepts conveyed in Figure \ref{Fig:MultiComps} can help with this procedure.  This produces the first and second moment of the subsystem, $E[F_S(t)]\equiv G_S(t)$ and $E[(F_S(t))^2]$.
\item Convert $E[(F_S(t))^2]$ to $\alpha_S(t)$ using equation \ref{Eq:prec}.
\item Finally, set the subsystem's prior to be
\[ F_S(t) \sim BSP(\alpha_S(t),G_S(t)).\] 
\end{enumerate}

The preceding modeling steps should be completed for each subsystem and iterated for each level of the system.  As stated in the literature review, the iteration of the modeling steps are similar to those in \cite{warr-2013-bayesian}, but significantly, in our methodology the correct precision is preserved and is informed by the data, additionally, right-censored data can be included in the analysis.

Once the posterior system BSP is found, an easy way to estimate its uncertainty is to use Monte Carlo simulation estimation.  This can be done by sampling realizations from the posterior system BSP.  We do this using the following procedure:
\begin{enumerate}
    \item For all jumps in the centering measure there exists a jump in the underlying L\'evy process, which we denote $S_k$ for $k\ in \{1,2,\ldots,n \}$.
    \item From Equation \ref{eq:jumps} in Appendix \ref{apx:changes}, each $S_k$ is related to a beta distribution as follows: \[ \exp\{-S_k\} \sim \text{Beta}\left( \alpha(t_k) (1-G(t_{k})) , \alpha(t_k) (G(t_k)-G(t_{k-1}))\right); \]
    randomly draw this quantity for each $k$.
    \item For any time $t$ there exist some largest $m$ such that $t_m \leq t$.  Then the randomly drawn CDF at time $t$ is defined by  
    \[ 1 - \prod^{m}_{i=1} \exp\{\-S_i\}.\]
\end{enumerate}
Sampling from the posterior in this manner can provide uncertainty estimates for the true system CDF.

\subsection{Priors for Components}
This section provides a brief discussion on how to quantify prior information in this model.  The priors are BSPs defined on the components, with a centering measure and a precision.  Although a precision set to zero implies no prior information, it should be avoided when possible.  If any prior information is available it should be incorporated into the analysis.  

A prior is given for a component's unknown CDF, which we denote $F(t)$. Again, $F(t)$ is a random CDF which models the true time-to-failure CDF for that component. Because the implementation of this model deals only with discrete CDFs, the centering measure in the BSP prior for each component will also be discrete. 

A straightforward method to obtain a prior for $F$ is to fit a DP to the modeler's prior information (i.e., $\alpha(t)$ is a constant function with respect to $t$).  To do this a modeler chooses $l$ time points that can be quantified in a manner described below.  Let $t_0 \equiv 0$ and $t_0 < t_1 < \ldots < t_l < \infty$.  These time points can be arbitrary, but they should be chosen such that the modeler has some belief of the value of $F(t_i)$, for $i \in \{1,2,\ldots,l\}$.  Once the belief of $F$ is quantified at those time points they should be codified in $G(t_i)$.  Then a precision must be selected.  

The precision determines the certainty (or lack thereof) for the belief of $F(t_i)$. Thus, if a modeler is quite certain (\textit{a priori}) that $G(t)$ closely approximates $F(t)$, the prior precision parameter should be large.  Conversely, if the a modeler is uncertain that $G(t)$ is a good approximation of $F(t)$ the prior precision parameter should be small. The DP's precision parameter can be interpreted as the amount of information contained in the prior and is an example of a data augmented prior \citep{christensen2010bayesian}.  Therefore, if the precision is equal to 10, the prior DP has the same posterior influence as 10 observations.  

Although the DP prior is not as flexible as a BSP, it is  straightforward to implement.  Clearly it would be possible to use a non-constant precision parameter and have a BSP prior.  That approach would be more complex to implement, but might more accurately quantify prior information.

\subsection{Computation Time} \label{sec:comp_time}

We now explore the computation time of the aforementioned methodology, specifically, the order of complexity of the sample size ($N$) and the number of components in the system. We focused these computations for parallel systems, but similar results hold for systems in series. 

The data were generated as follows: for each component, we simulated $N$ observations from a Weibull(2,40) distribution. 
Then we introduced right censoring in approximately 10\% of the data, with censoring times occurring only in the right tail of the distribution.
We included no full system data and only weak prior information to minimize any confounding of the simulation results.  The prior centering measure is set at the true value of the CDF on the support points at $0,5,\ldots,95, 100$ with a low fixed precision of 0.2.

\begin{figure}[!ht]
    \centering
    \includegraphics[width=\textwidth]{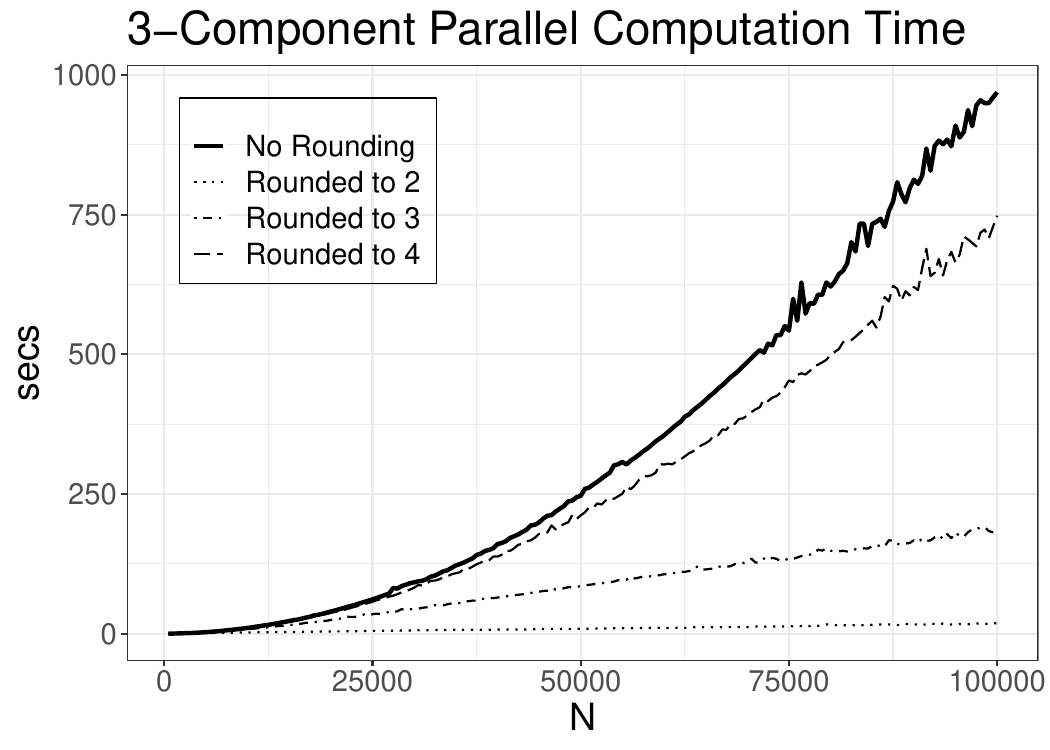}
    \caption{A plot (solid line) showing the computation required for a 3-component parallel system as $N$ increases.  When there is no rounding and each data point is unique, the algorithm to calculate the posterior is approximately $O(N^2)$. If the observations are rounded, the other lines show the computational effect; which have significantly reduced computation time.}
    \label{comp_n_round}
\end{figure}

The computation time included the time required for calculating the posterior of all components and the full system. Table \ref{comp_N} and Figure \ref{comp_n_round} show the effect of increasing $N$ in a 3-component parallel system on computation time. The full posterior will end up having about $3 N$ support points. When $N$ is large (e.g. 100,000), this results in a very large posterior support, which slows computation time. The order of complexity roughly appears to be $O(N^2)$. However, having several thousand support points with 5+ decimal accuracy may not be necessary. To drastically speed up the computation time, the data were also rounded. The effect of rounding on computation time is shown visually in Figure \ref{comp_n_round}. 

\begin{table}[ht]
\centering
\small
\begin{tabular}{l|llll}
N        & Round to 2 & Round to 3 & Round to 4 & No Rounding \\\hline
$10^2$      & 0.03 sec      & 0.03 sec      & 0.03 sec & 0.04 sec       \\
$10^3$      & 0.14 sec      & 0.16 sec      & 0.15 sec & 0.19 sec       \\
$10^4$     & 2.04 sec      & 7.95 sec      & 10.03 sec & 10.21 sec     \\
$10^5$   & 18.2 sec   & 3.2 min    & 13.1 min          & 16.2 min       \\
$10^6$   & 3.6 min    & 45.1 min   & 5.8 hrs            & 28.1 hrs    
\end{tabular}

\caption{Computation times for a 3-component system in parallel. Note that if there is no rounding of the data (or all observed times are unique), then computationally, increasing $N$ appears to be approximately $O(N^2)$. 
}
\label{comp_N}
\end{table}

\begin{figure}[!ht]
        \centering
        \begin{subfigure}[b]{0.49\textwidth}
                \includegraphics[width=\textwidth]{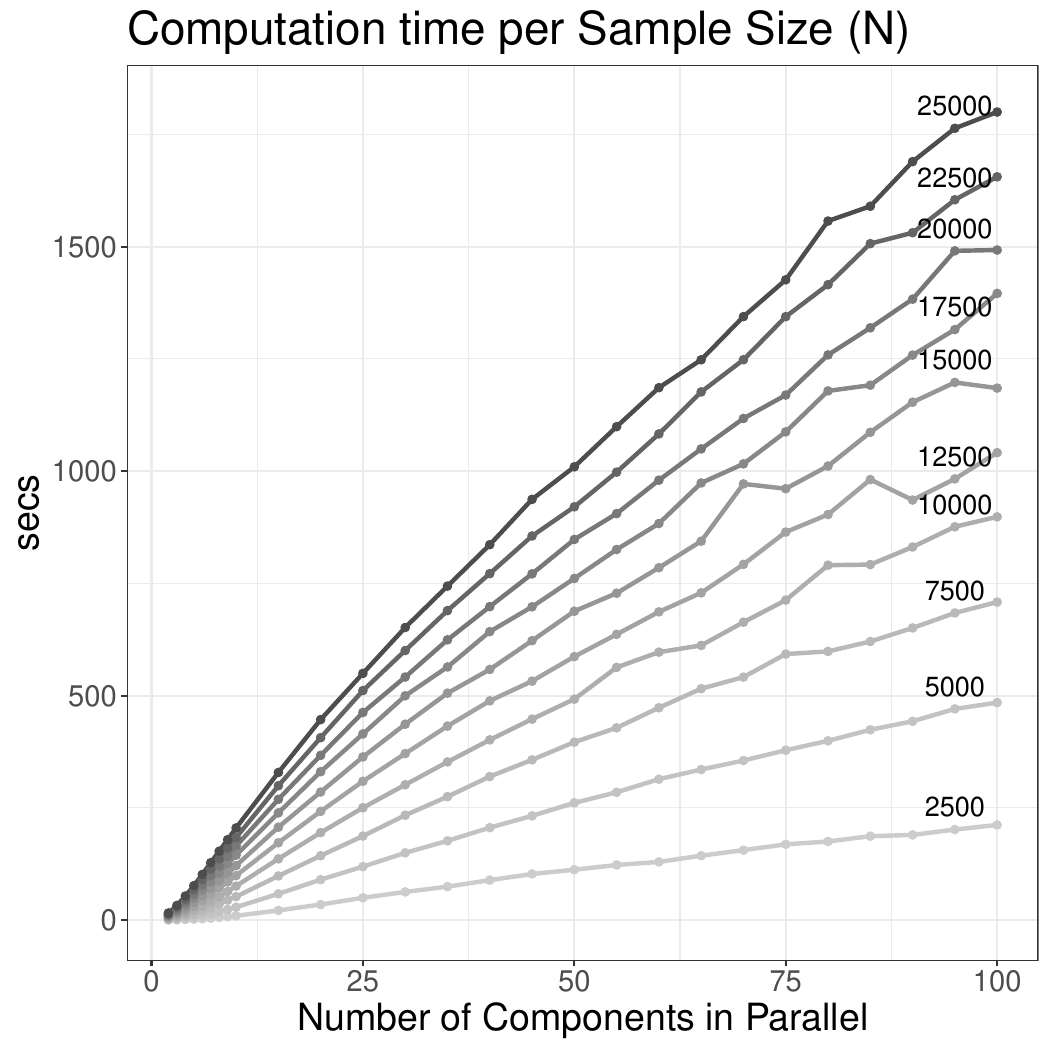}
                \caption{The lines represent the sample size for each component.}
                \label{comp_costs_C}
        \end{subfigure}%
        ~ 
        \begin{subfigure}[b]{0.49\textwidth}
                \includegraphics[width=\textwidth]{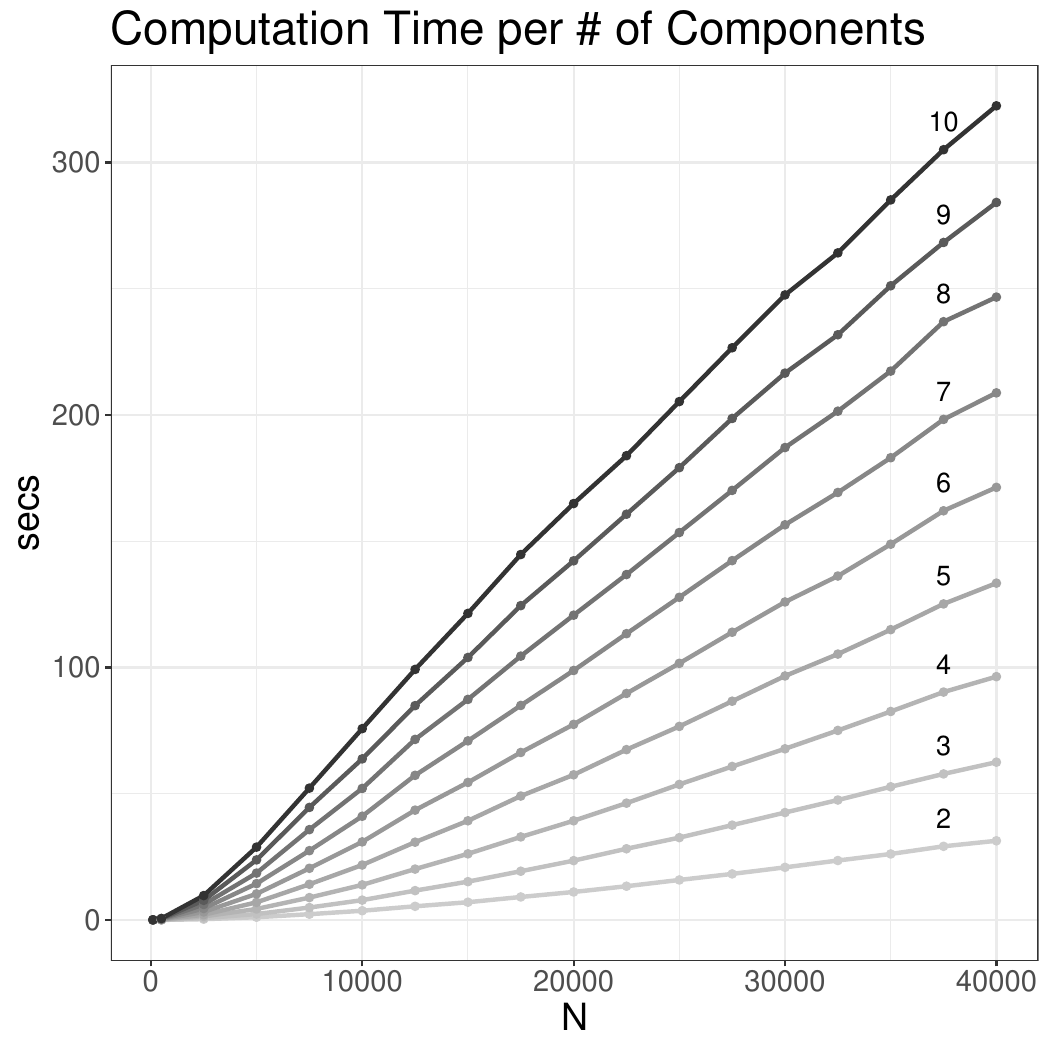}
                \caption{The lines represent the number of components in the system.}
                \label{comp_costs_N}
        \end{subfigure}
        \caption{Computational times to get a posterior after increasing the number of components in parallel and the sample size. The data were rounded to 3 decimal places before analysis. Note the slight concave shape in (a) and the slight convex shape in (b).}
        \label{comp_costs}
\end{figure}

A decrease in computation time is achieved by reducing the number of possible support points. In this example, rounding to 4 decimal places resulted in a 20.7\% decrease in posterior support points, rounding to 3 resulted in a 76.4\% decrease, and rounding to 2 resulted in a 97.1\% decrease. Despite these decreases, the full posteriors did not have sparse support: they had 213,643, 63,712 and 7,818 support points, respectively. We did not experience a drastic decrease in accuracy after rounding the data for large $N$. When comparing the posterior centering measure to the true CDF, rounding the data to 3 decimals had a mean absolute error (MAE) of .000495 compared to a MAE of .000460 in the case of no rounding. 

Realistically, data are usually observed with a finite level of precision. Thus, a very large data set will likely have duplicate times if the precision is not too large. With real data, one can expect computation times to be less than $O(N^2)$ but greater than $O(N)$. If even faster times are desired, the modeler can round the data before calculating the posterior. By doing so, the total computation time is drastically reduced for a small penalty in accuracy, and the methodology is much more scalable (it has an order of complexity closer to $O(N)$).

We completed an additional study exploring the effect of increasing the number of components on the system posterior computation time. We used the same procedure as outlined earlier in this section to generate the data. We rounded the data to 3 decimal places and measured the time required to obtain a full posterior. Figure \ref{comp_costs} shows the effect on computation time after adjusting both $N$ and the number of components. From the concave nature of Figure \ref{comp_costs_C}, we can see that increasing the number of components seems to scale at some order at or slightly less than $O(N)$. If the data were not rounded before analysis, increasing the number of components behaves much closer to $O(N)$. This is in contrast with the convex shape in Figure \ref{comp_costs_N}, which suggests that regardless of the number of components, $N$ seems to scale slightly greater than $O(N)$. 

\section{Simulation Study}
We propose a simulation study which explores the performance of our proposed model against a traditional parametric approach. We evaluate 3 different models on simulated data: our BNP model, a correctly specified parametric model, and a misspecified parametric model. We explore the resulting computation time, mean absolute error, and bias of the estimated system CDF against the true CDF. 

\begin{figure}[!ht]
\begin{center}
\includegraphics[width=5in]{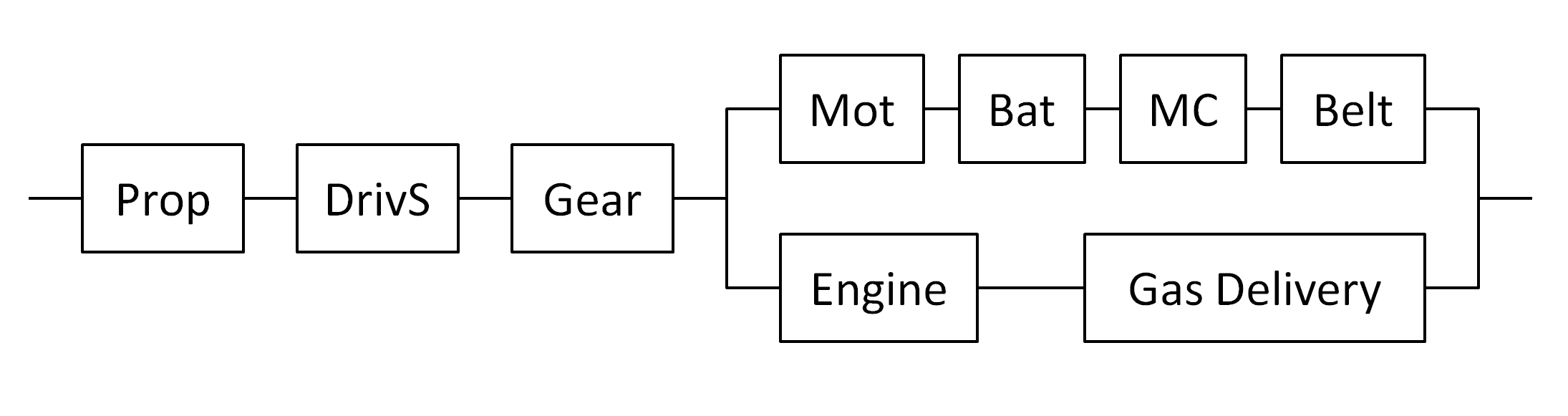}
\caption{The reliability block diagram used to generate data for the simulation study.}
\label{Fig:RelBlock}
\end{center}
\end{figure}

We generate the data for the simulation based on a hybrid-electric aircraft known as SHERPA, which was under development at the Air Force Institute of Technology. The reliability goal we consider is simply whether the aircraft can fly. This aircraft uses a dual propulsion system (electric motor and gas engine) for takeoff, while it can remain in flight if either one is operable.  Table \ref{tab:CompList} lists the components of the aircraft's propulsion system. The interrelationship of components is shown by the reliability block diagram in Figure \ref{Fig:RelBlock}. The three common components on the left are in series, since a failure of any one results in loss of propulsion. The two parallel branches to the right indicate that there is propulsion if either the gas engine or electric motor is functional.  

\begin{table} \centering
\caption{Simulation study component list}
\label{tab:CompList}
\begin{tabular}{ l l l }
  Common Parts & Electric Propulsion & Gasoline Propulsion  \\ \hline
  Propeller & Motor & Engine  \\
  Drive shaft & Batteries & Gas Delivery  \\
  Gearing & Motor controller &   \\
  & Serpentine belt & \\ \hline
\end{tabular}
\end{table}

Failure times for the common parts were simulated with an exponential distribution with rate parameters 1/2, 1/3 and 1, respectively (from left to right in Figure \ref{Fig:RelBlock}). Failure times for the electric propulsion were generated with a lognormal distribution with parameters (1,1), (1/5,1/3), (5/2, 1/3), and (1,1/2) respectively. The gas components were generated using gamma distributions with shape and rate parameters (1/4,1/4) and (1/10, 1/10), respectively. Data were generated for the electric, gas, and common subsystems, as well as the full system. We generated 30 observations from the nine components, the three subsystems, and the entire system.  Right censoring was introduced, which resulted in approximately 15\% of the data being censored. 

We compared the performance of our method against a Bayesian parametric approach on the aircraft's reliability. The metrics for comparison are the mean absolute error (MAE) and the bias of 3 different model choices of the full system CDF. 
We estimated the full system CDF using 1) our proposed method, 2) a correctly specified parametric model, and 3) an incorrectly specified parametric model. The incorrectly specified model assumes the common parts are distributed lognormal, and the electric/gas parts are distributed Weibull. Real component data often does not nicely match known distributions. Thus, we assume mild misspecification in the parametric model choice. No prior information is included for the components in our model. We used reference priors in the parametric models. The priors for the lognormal $\mu$ parameters are N(0, 100) and all other parameters have a prior of Gamma(.1, .1). 

Posterior draws were simulated using a Gibbs sampler in the parametric models. The parameters (15 in the correct model, 18 in the incorrect model) were assessed for convergence and mixing. The posterior mean for each parameter was then used to construct a CDF. While the run-time for the full posterior of the BSP method is under 2 seconds on a personal computer, the run-time for 10,000 MCMC draws for each parametric model is about 45 mins. Even in a model with just 9 components, the parametric method took significantly longer than our proposed method. Posterior inference using parametric methods quickly becomes computationally expensive as system complexity increases. 

\begin{table}
\centering
\caption{Estimation results for the three model specifications.}
\begin{tabular}{r|ccc}
& & Correct & Misspecified \\
& Nonparametric &  Parametric &  Parametric \\
\hline
MAE & 0.0311    & 0.0179      & 0.0391    \\
Average Bias & 0.0021 & 0.0048 & 0.0280 \\
\hline
\end{tabular}
\label{mae_tab}
\end{table}

\begin{figure}
    \includegraphics[width=\textwidth]{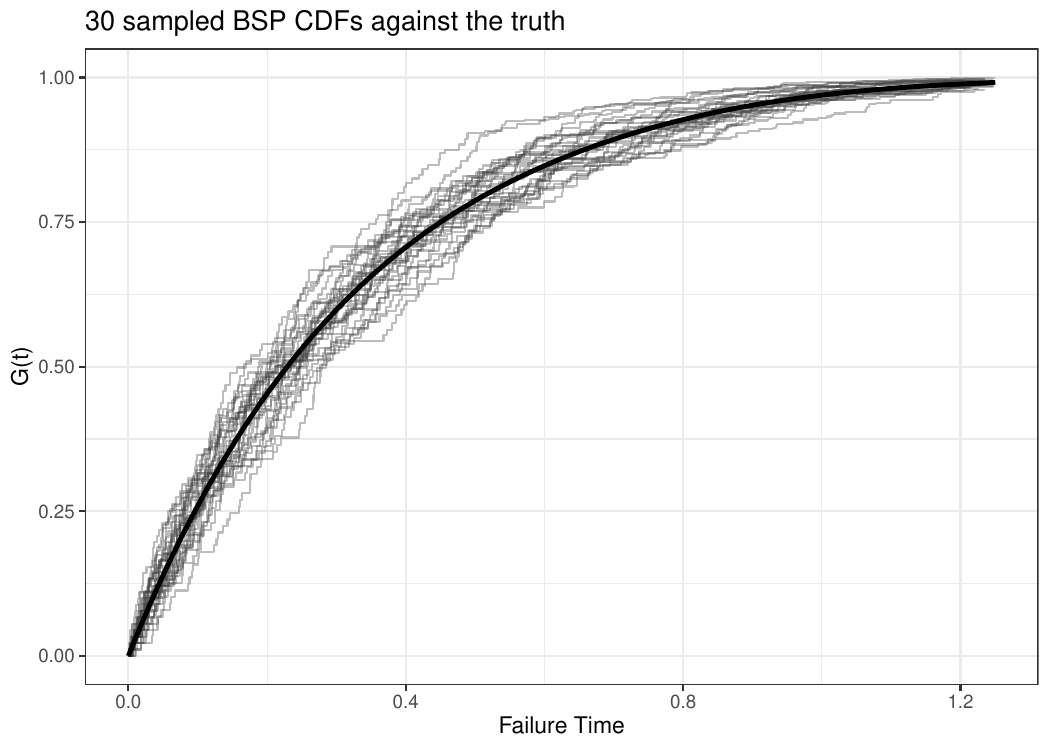}
    \caption{Simulated BSP posterior system centering measures from 30 datasets (in gray), with the true system CDF in black. The datasets were generated with the same parameters.}
    \label{bsp_cdfs}
\end{figure}

\begin{figure}
    \includegraphics[width=\textwidth]{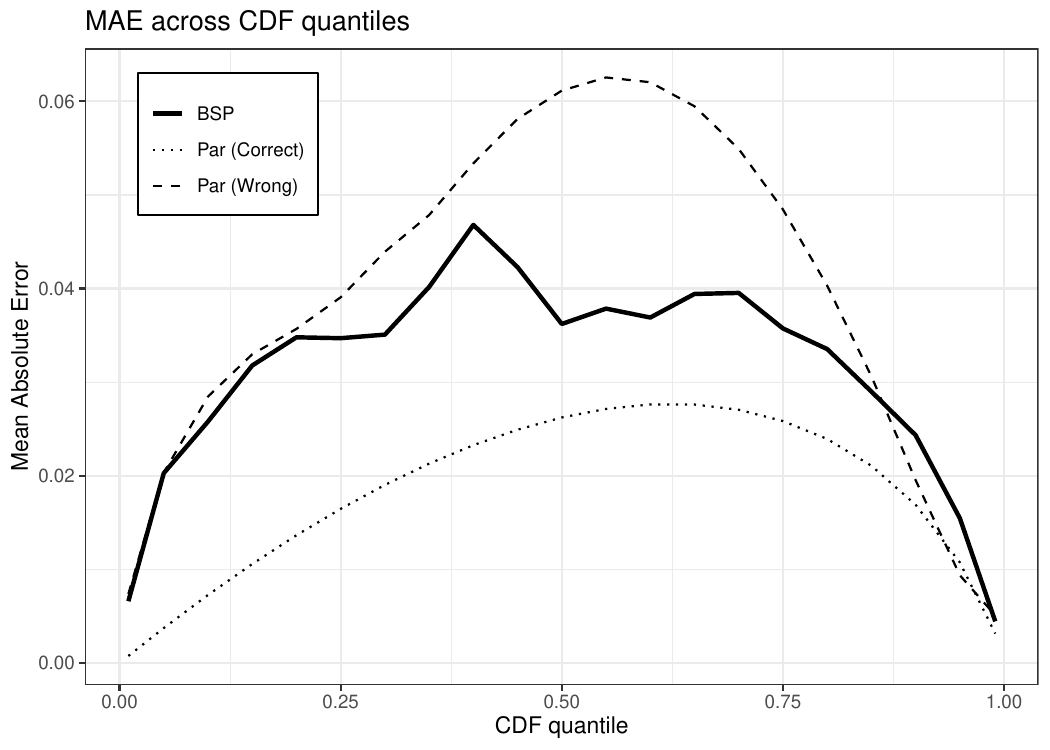}
    \caption{Mean absolute error across quantiles of the true CDF. }
    \label{mae_quantiles}
\end{figure}

We simulated 30 different sets of data, built the three different models, and extracted only the full system CDF (with its uncertainty). Figure \ref{bsp_cdfs} shows 30 full system CDFs from the BSP method plotted against the truth. We compared the estimated CDF against the true CDF values at the $0.01, 0.05, 0.1, ... , 0.95, \text{ and } 0.99$ quantiles. Table \ref{mae_tab} shows the aggregated the MAE across the three different models. While the correctly specified parametric model performs the best, the BSP method outperforms the parametric misspecified model. The errors are not the same across the entire CDF. Figure \ref{mae_quantiles} shows the MAE across the true CDF quantiles. Unsurprisingly, the correctly parameterized method does the best across most of the support, but even for mild misspecification, the BSP method performs better. 


\begin{figure}
    \includegraphics[width=\textwidth]{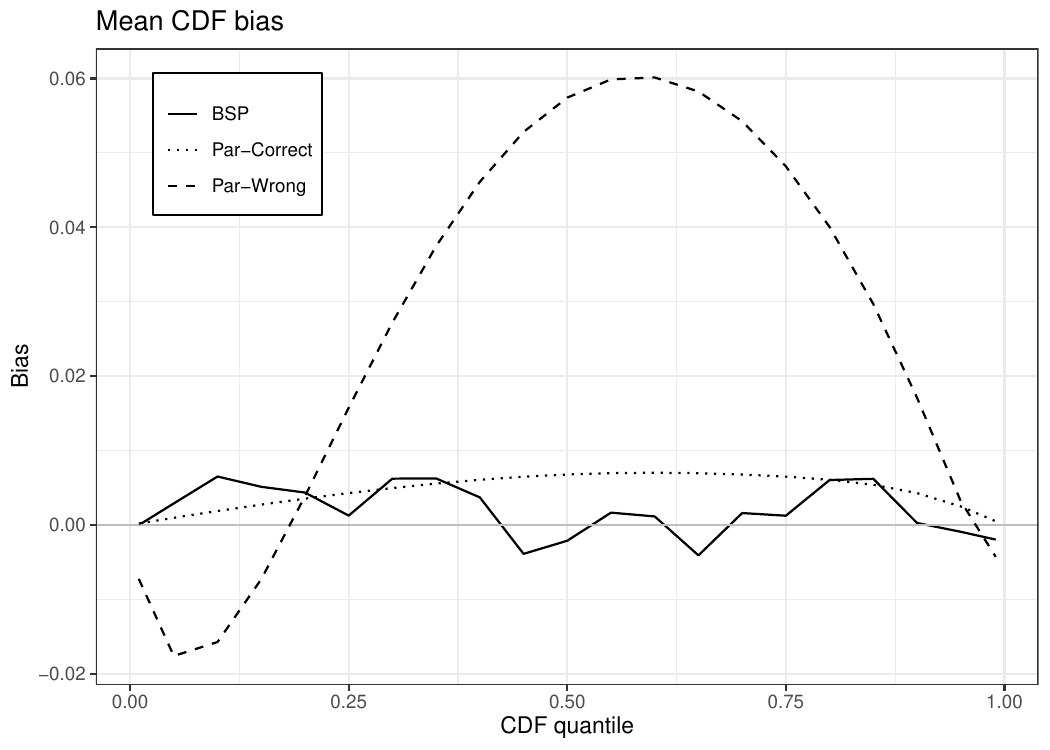}
    \caption{Bias for the 3 different model specifications. Note the consequences on the bias after misspecifying the parametric model.}
\label{sim_bias}
\end{figure}

The bias of the three models is shown in Figure \ref{sim_bias}. The misspecified parametric model is biased much higher than would be desirable. Due to the constraints of an incorrect model, the estimated CDF has a difficult time fitting the true curve. The large bias in the misspecified model is likely the cause of the increased MAE in Figure \ref{mae_quantiles}. A moderate misspecification and/or violation of parametric assumptions increases the risk of incorrectly estimating CDFs. The correctly parameterized model is slightly biased, which may be a result of inherit bias in the methodology, or just bias from the priors. Our method was able to remain roughly unbiased across the CDF. The reason the proposed method had a somewhat high MAE is not because the CDF estimate is inaccurate, but because the centering measure estimator for the true system CDF has a higher variance.  Specifying the true model is often unrealistic, thus the proposed method can often be a safer option because we are trading an increase in variability for a smaller bias.  Even if a traditional parametric model is preferred, our method can serve as a quick check that the parametric model is behaving appropriately. 

\section{Application}

A manufacturing company (which remains nameless) was interested in estimating the reliability of a new product.  This product is comprised of six components in series.  We demonstrate the flexibility and speed of our model for this new system.

There were a couple sources of data to estimate the system reliability.  The first data source was from similar components from previous systems.  The amount of data available from these previous systems was impressive.  However, the components do not identically map from the older systems to the new system.  Components in the new system were considered to be worse than, about the same as, and better than the old components.  Thus a straightforward way to generate the priors for the new system components was to adjust the centering measure according to engineering judgement.  If the new component was considered to be worse than the old, then we would fit a CDF to the old data and perform some time-scaling.  For example, consider an older component to which we fit a CDF to its failure times (this can be done using 1 minus the Kaplan-Meier estimator).  We call this fit $G_0(t)$.  If the engineers estimate that the new component would only perform 80\% as well as the older one then the centering measure for the component's prior is set to $G(t) = G_0(t/0.8)$.

To complete the prior for a component, a precision needs to be selected.  This value is also set using engineering judgment, and represents the prior's influence (in number of observations) in the component posterior.  
The other data are also at the component level, and are field test data of the new components.     

The data for this product is proprietary and cannot be released.  To demonstrate the concepts of our method, we obfuscated the data to protect sensitive corporate information.
For the older component data we re-sampled 2000 observations from the original data (each component had more than 2000 observations).  Then using that data an initial centering measure was set.  The engineers then determined any scaling that needed to happen.  Of the six components, one was determined to be twice as good as the old, three were half as good as the old, and the other two remained the same.  

\begin{figure}[!ht]
    \centering
    \includegraphics[width=\textwidth]{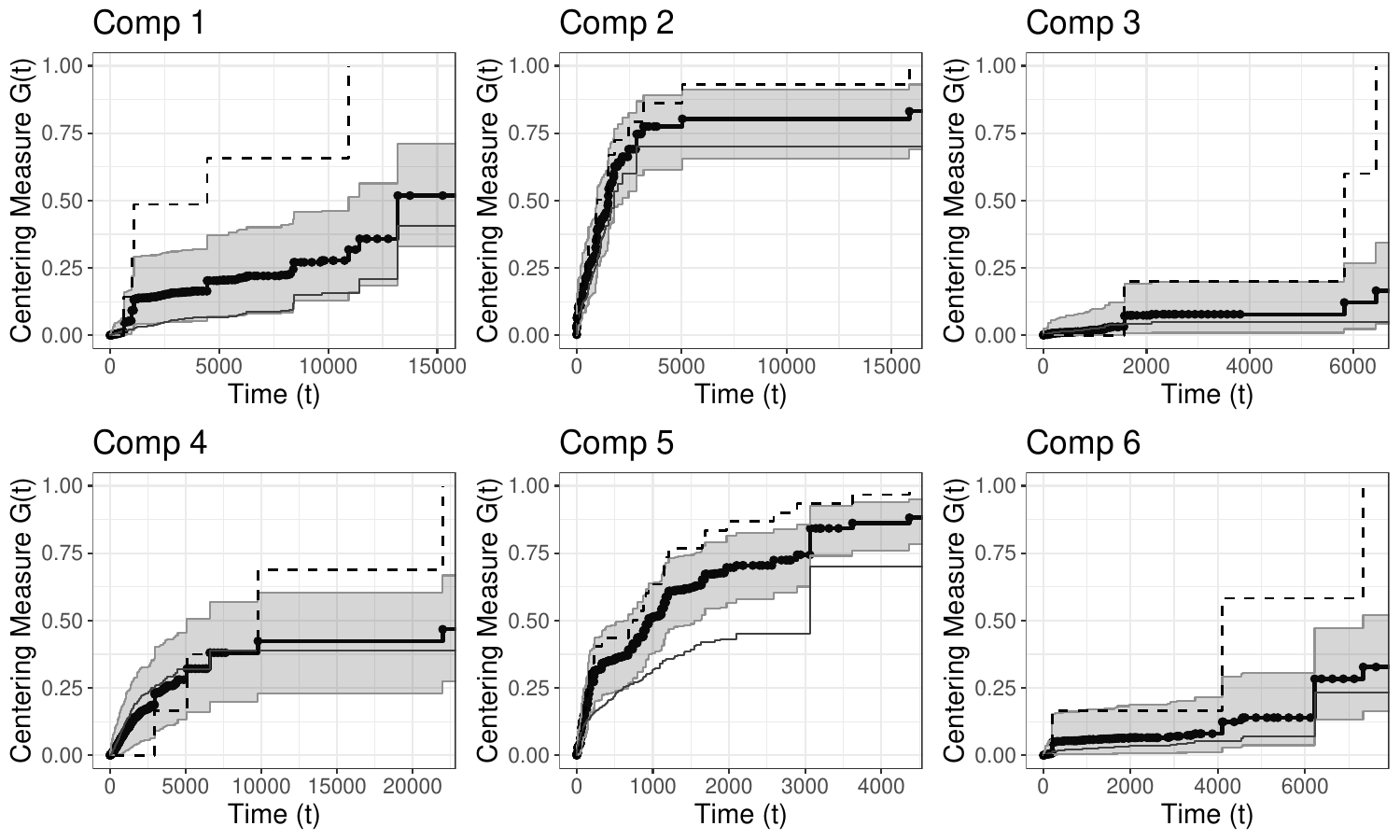}
    \caption{Posterior distributions of each of the 6 components. The thick black line and shaded region represent the posterior mean and 95\% credible interval for the CDF. The dashed line indicates the Kaplan-Meier estimate from the field data and the thin solid line indicates the mean of the prior. }
    \label{app_comps}
\end{figure}

The actual component (field) data were disguised by fitting a Weibull to each and then sampling the same number of observations (and censoring).  Table \ref{app_params} shows the Weibull model parameters, that generated the analyzed data, and the number of samples for each component.  The Weibull parameterization is defined by $E(X) = (\text{scale})\Gamma(1+\text{1/shape})$.  Some plots of the priors and posteriors of the components are shown in Figure \ref{app_comps}.  

\begin{table}[ht]
\centering
\caption{Sample sizes and Weibull parameters for the field test data for the 6 components.}
\begin{tabular}{l|llllll}
 Component   & 1     & 2     & 3     & 4     & 5     & 6     \\\hline
scale   & 7323  & 1601  & 12313 & 8294  & 1027  & 11908 \\
1/shape & 0.852 & 1.583 & 1.428 & 1.165 & 1.391 & 1.596 \\\hline
Total N       & 8 & 23 & 8 & 8 & 36 & 7 \\
N censored & 3 & 5 & 5 & 4 & 4 & 4 \\\hline
\end{tabular}
\label{app_params}
\end{table}

\begin{figure}[!ht]
    \centering
    \includegraphics[width=\textwidth]{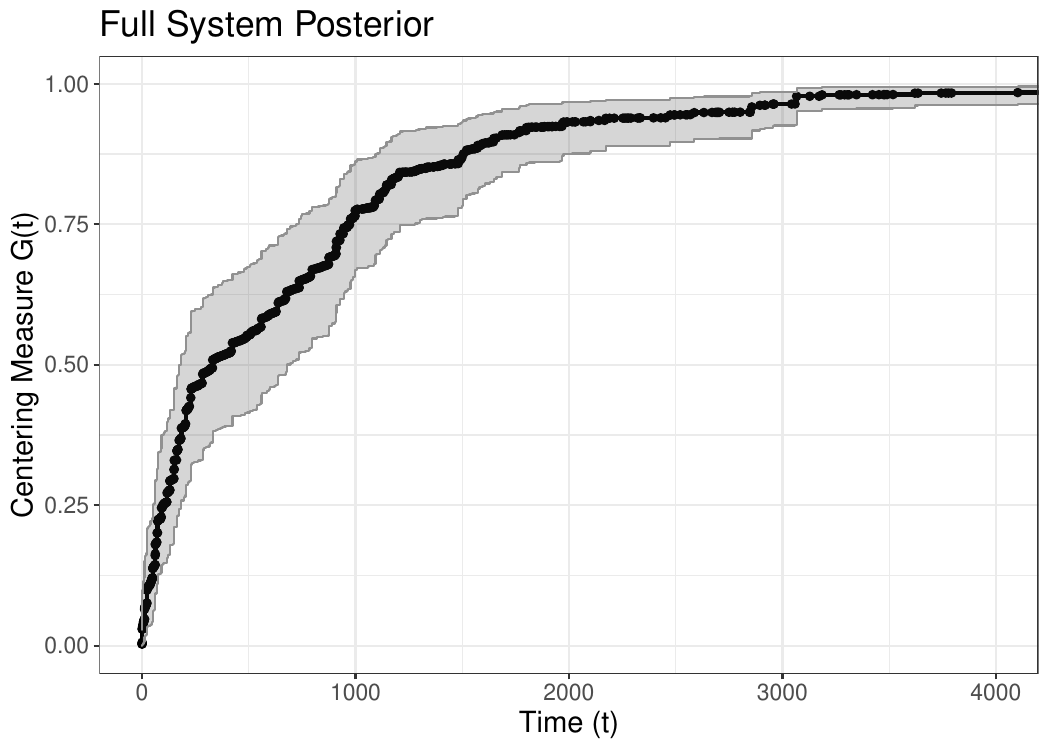}
    \caption{Full system posterior distribution with 95\% point-wise credible intervals in the shaded region.}
    \label{app_fullsys}
\end{figure}

The six components were combined into a single system in series. This data was used to update the prior beliefs and generate a posterior. The full system is shown in Figure \ref{app_fullsys}. Summary statistics of the full 6 component system are shown in Table \ref{app_fullsys_summary}. The total time required to simulate the data, calculate the posterior, generate Monte Carlo samples for the credible intervals, and make the plots was approximately 1.88 seconds. 

\begin{table}[!ht]
\caption{Posterior summary statistics of system lifetime}
\centering
\begin{tabular}{r r r r r r r r r}
    mean & sd & 5\% & 10\% & 25\% & 50\% & 75\% & 90\% & 95\% \\
    \hline
   716.7 & 1127.5 & 11.5 & 23.4 & 96.5 & 330.0 & 972.0 & 1646.0 & 2849.0 \\
    \hline
\end{tabular}
\label{app_fullsys_summary}
\end{table}

This information can be used to forecast warranty costs and find opportunities for reliability growth.  And, as new field data arrive, these results can be updated almost instantly.

\section{Discussion}

The methods proposed in this paper developed a Bayesian system reliability model using system data augmented with prior information from subsystems and components.  Data from any one of those sources could be absent and the method would still work well, but it would produce estimates with wider uncertainty.  Computation of the posterior estimates is very fast in relation to other methods, particularly Bayesian parametric methods that rely on MCMC computations.

There are some areas that merit further exploration.  One is the primary model assumption that components are independent.  If the independence assumption is violated the model framework loses accuracy (depending on the amount of dependence that exists).  One possible way to build that into the model would be to use copulas.  Copula models would be a natural extension to incorporate dependence into the model.  For a popular introduction to copulas see \cite{nelsen2007introduction}.

Another obvious model extension that could be explored is incorporating other types of reliability data such as interval censored data.  However, since there are no known conjugate models for interval censored data, this would introduce a fairly severe computational cost.  

In summary, the proposed model is well suited for large-scale reliability applications that must be run multiple times.  If data are being reported on a continual basis this model is would be ideal, with the minimal assumptions and fast computation times.  It could also be used to help validate parametric models.

\section{Acknowledgments}
The authors thank the B61 Life Extension Program at Los Alamos National Laboratory for providing funding which made a part of this work possible.  We also thank Brandon Greenwell for his work and insights on this project.

\bibliographystyle{elsarticle-harv}

\newpage
\begin{appendices}

\section{\hspace{-19pt}: Changes to the BSP}
\label{apx:changes}

As mentioned in the article we made some minor adjustments to the beta-Stacy process to suit our needs.  \\

\noindent {\bf Change 1}\\
The first change we make to the BSP comes near the bottom of page 3 in \citet{walker-1997-beta}.  Their equation is
\[ 1-\exp(-S_k) \sim \mathcal{B}\left( c(t_k) G\{t_k\}, c(t_k) G[t_k,\infty)\right). \]
We change it to
\begin{equation} \label{eq:jumps}
 1-\exp(-S_k) \sim \mathcal{B}\left( c(t_k) G\{t_k\}, c(t_k) G(t_k,\infty)\right).     
\end{equation}
According to our calculations this change is needed prove $E[F(t)]=G(t)$ (Theorem \ref{thm:mean}).  For a proof of our Theorem \ref{thm:mean} see Appendix \ref{proof:mean}.   We can justify this change theoretically, by replacing their Equation (4) with an alternate definition of the cumulative hazard function.  They cite \cite[pg. 1260]{gill1990survey} to define the cumulative hazard function, but differing definitions exist which support our change, see \citet[pg. 459]{aalen2008survival} and \citet[pg. 92]{andersen2012statistical}.\\

\noindent {\bf Change 2}\\
The other change we make to the BSP is the formula for the posterior precision parameter.  On page 1775 of \cite{walker-1997-beta} they have
\[ c^*(t) = \frac{c(t)G[t,\infty) + M(t) -N\{t\} }{G^*[t,\infty)}.\]
We change this to
\[ c^*(t) = \frac{c(t)G[t,\infty) + M(t) -N\{t\} }{G^*(t,\infty)}.\]
This change is reflected in our Theorem \ref{Eq:postprec} and it allows the BSP to be internally consistent.  For example, if no data were collected, the posterior should be equal to the prior.  This change ensures that this will occur (see Appendix \ref{apx:post} for a demonstration).  

\newpage
\section{\hspace{-19pt}: Simple BSP Posterior Examples}
\label{apx:post}

To demonstrate Equations \ref{Eq:postbase} and \ref{Eq:postprec} consider the CDF
\begin{equation*}
H(t) = 
\begin{cases} 
0 &\mbox{for } t < 1 \\ 
1/3 &\mbox{for } 1 \leq t < 2\\
2/3 &\mbox{for } 2 \leq t < 3\\
1 &\mbox{for } t \geq 3. 
\end{cases}
\end{equation*}
Let $H(t)$ be the prior and suppose no other data are collected.  Then the posterior $G^*(t)$ should be equal to the prior $H(t)$.
\begin{align*} G^*(1) &= 1-\prod_{i=1}^1\left( 1-\frac{\alpha(t_i)\left(H(t_i)-H(t_i-)\right)+J(t_i) }{\alpha(t_i)\left(H(t_i)-H(t_i-)\right)+J(t_i)} \right) \\ &= 1-\left(1-\frac{\alpha(1)(1/3-0)+0}{\alpha(1)(1-0)+0}\right)=1/3=H(1), \end{align*}
\[G^*(2) = 1-\left(\frac{2}{3}\right)\left(1-\frac{\alpha(2)(2/3-1/3)+0}{\alpha(2)(1-1/3)+0}\right)=2/3=H(2),\]
and
\[G^*(3) = 1-\left(\frac{2}{3}\right)\left(\frac{1}{3}\right)\left(1-\frac{\alpha(3)(1-2/3)+0}{\alpha(3)(1-2/3)+0}\right)=1=H(3). \]
The posterior precision should also be equal to the prior precision and we obtain
\[ \alpha^*(1) = \frac{\alpha(1)1+0-0}{1} = \alpha(1), \]
\[ \alpha^*(2) = \frac{\alpha(2)2/3+0-0}{2/3} = \alpha(2), \]
and 
\[ \alpha^*(3) = \frac{\alpha(3)1/3+0-0}{1/3} = \alpha(3). \]
This simple example shows a few of the mechanics of the conjugacy of the BSP.

Additionally, the converse scenario also produces reasonable results.  Assume no prior information is available (i.e., $\alpha(t)\equiv0$) and there are three failures which occur at time points 1,2, and 3, then
\[G^*(1) = 1-\prod_{i=1}^1\left( 1-\frac{0\left(H(t_i)-H(t_i-)\right)+J(t_i) }{0\left(H(t_i)-H(t_i-)\right)+J(t_i)} \right) = 1-\left(1-\frac{1}{3}\right)=1/3, \]
\[G^*(2) = 1-\left(\frac{2}{3}\right)\left(1-\frac{1}{2}\right)=2/3, \]
and 
\[G^*(3) = 1-\left(\frac{2}{3}\right)\left(\frac{1}{3}\right)\left(1-\frac{1}{1}\right)=1. \]
Which is just the empirical CDF of the data.
For the precision parameter we have
\[ \alpha^*(1) = \frac{0\left(1-G(1)\right)+3-1}{1-1/3} = 3, \]
\[ \alpha^*(2) = \frac{0\left(1-G(2)\right)+2-1}{1-2/3} = 3, \]
and 
\[ \alpha^*(3-) = \frac{0\left(1-G(3-)\right)+1-0}{1-2/3} = 3. \]
Which is a constant of $3$ so the posterior is a DP as one would expect.  

\newpage
\section{\hspace{-19pt}: Proofs and Derivations}
\label{apx:precision}

This appendix contains the proofs for the theorems in this article.

\subsection{Proof of Theorem \ref{thm:mean}} \label{proof:mean}
We start by considering the locations and heights of the jumps of the defining L\'evy process.  The locations are at $t_1,\ldots,t_n$ and the jumps are of height $S_1,\ldots,S_n$.  The CDF is defined as 
\[F(t) = 1-\prod_{i=1}^{m} \exp(-S_i),\]
where $m$ is defined as the largest index such that $t_m < t$.
From Equation \ref{eq:jumps} we know 
\[ \exp(-S_i) \sim \text{Beta}\Big( \alpha(t_i) \big(1-G(t_i)\big), \alpha(t_i) \big(G(t_i)-G(t_i-)\big) \Big), \]
thus $E[\exp(-S_i)] = \big(1-G(t_i)\big)/\big(1-G(t_i-)\big)$.
Since the jumps are independent we have 
\begin{align*}
    E[F(t)] &= 1-\prod_{i=1}^{m} E[\exp(-S_i)] \\
    &=1 - \frac{1-G(t_1)}{1-G(t_1-)}\frac{1-G(t_2)}{1-G(t_2-)}\cdots \frac{1-G(t_m)}{1-G(t_m-)}\\
    &=1 - \frac{1-G(t_1)}{1-G(t_0)}\frac{1-G(t_2)}{1-G(t_1)}\cdots \frac{1-G(t_m)}{1-G(t_{m-1})}\\
    &=1 - \frac{1-G(t_m)}{1-G(t_0)}\\
    &=1 - \frac{1-G(t)}{1-G(0)} = 1 - \frac{1-G(t)}{1-0} = G(t) \qed
\end{align*} 

\subsection{Proof of Theorem \ref{thm:2ndmoment}} \label{proof:2nd}

From the previous proof we can start with 
\begin{align*}
    E[(1-F(t))^2] &= \prod_{i=1}^{m} E[(\exp(-S_i))^2] \\
    E[(F(t))^2] - 2 G(t) +1 &= \left(\prod_{i=1}^{m} E[(\exp(-S_i))^2]\right).
\end{align*} 
But the second moment of a Beta$(a,b)$ is
\[\frac{a(a+1)}{(a+b)(a+b+1)}.\]
Thus
\begin{align*}
   \prod_{i=1}^{m}  E[(\exp(-S_i))^2] =   
   &\prod_{i=1}^{m} \frac{\alpha(t_i)(1-G(t_i)) [\alpha(t_i)(1-G(t_i))+1]}{\alpha(t_i)(1-G(t_i-))[\alpha(t_i)(1-G(t_i-))+1]}\\
  &\prod_{i=1}^{m} \frac{(1-G(t_i)) [\alpha(t_i)(1-G(t_i))+1]}{(1-G(t_i-))[\alpha(t_i)(1-G(t_i-))+1]}.
\end{align*} 
Putting this together we have
\begin{equation*}
    E[(F(t))^2] = \left(\prod_{i=1}^{m} \frac{(1-G(t_i)) [\alpha(t_i)(1-G(t_i))+1]}{(1-G(t_i-))[\alpha(t_i)(1-G(t_i-))+1]} \right) -1+ 2 G(t) \qed
\end{equation*}

\subsection{Proof of Theorem \ref{thm:prec}} \label{proof:prec}

\noindent To prove Theorem \ref{thm:prec} we start with the results of Theorem \ref{thm:2ndmoment} and drop the $S$ subscripts on $F$ and $G$ to reduce the notational clutter.  Thus, for any $t>0$ there is a corresponding $t_m$, then we have

\begin{align*}
E[(F(t_m))^2] = & \left( \prod_{i=0}^{m} \frac{(1-G(t_i))[\alpha(t_i)(1-G(t_i))+1]}{(1-G(t_i-))[\alpha(t_i)(1-G(t_i-))+1]} \right) -1 +2G(t_m) \\
 = & \left( \prod_{i=0}^{m-1} \frac{(1-G(t_i))[\alpha(t_i)(1-G(t_i))+1]}{(1-G(t_i-))[\alpha(t_i)(1-G(t_i-))+1]} \right) \times \\ & \left( \frac{(1-G(t_m))[\alpha(t_m)(1-G(t_m))+1]}{(1-G(t_m-))[\alpha(t_m)(1-G(t_m-))+1]} \right) -1 +2G(t_m) \\
 = & \left( E[(F(t_{m-1}))^2] + 1 - 2G(t_{m-1}) \right) \times \\ & \left( \frac{(1-G(t_m))[\alpha(t_m)(1-G(t_m))+1]}{(1-G(t_{m-1}))[\alpha(t_m)(1-G(t_{m-1}))+1]} \right) -1 +2G(t_m)  \\
\end{align*}
Since $G(t)$ is a right continuous discrete step function, $G(t_m-) = G(t_{m-1})$. We now use the above equation and solve for $\alpha(t_m)$ and use some of the notation from Equation \ref{Eq:prec}. 
\small
\begin{align*}
E[(&F(t_{m}))^2] + 1 - 2G(t_m) = \\
&\left( E[(F(t_{m-1}))^2] + 1 - 2G(t_{m-1}) \right) \left(\frac{(1 - G(t_{m}))[\alpha(t_m)(1 - G(t_{m})) + 1]}{(1 - G(t_{m-1}))[\alpha(t_m)(1 - G(t_{m-1})) + 1]} \right)  
\end{align*}
\begin{align*}
\big( E[(F(t_{m}&))^2] + 1 - 2G(t_m) \big) \left( (1 - G(t_{m-1}))[\alpha(t_m)(1 - G(t_{m-1})) + 1] \right) = \\
&\left( E[(F(t_{m-1}))^2] + 1 - 2G(t_{m-1}) \right) \left((1 - G(t_{m}))[\alpha(t_m)(1 - G(t_{m})) + 1] \right)  
\end{align*}
\begin{align*}
Numerator_2 \, [\alpha(t_m)(1 - G(t_{m-1})) + 1]  = 
Numerator_1 \, [\alpha(t_m)(1 - G(t_{m})) + 1] 
\end{align*}
\begin{align*}
Numerator_2 \, \alpha(t_m)(1 - G(t_{m-1}&)) + Numerator_2  = \\ &Numerator_1 \, \alpha(t_m)(1 - G(t_{m})) + Numerator_1 \end{align*}
\begin{align*}
\alpha(t_m) \big( Numerator_2 \, (1 - G(t_{m-1})) - Numerator_1 & \, (1 - G(t_{m})) \big)   = \\ & Numerator_1 - Numerator_2 
\end{align*}
\begin{align*}
\alpha(t_m) \big( Denominator_1 - Denominator_2 \big)    =  Numerator_1 - Numerator_2 
\end{align*}
\begin{align*}
\alpha(t_m) = \frac{ Numerator_1 - Numerator_2 } { Denominator_1 - Denominator_2 } \qed
\end{align*}
\normalsize

\end{appendices}
\end{document}